\documentclass[conference]{IEEEtran}
\IEEEoverridecommandlockouts
\usepackage{cite}
\usepackage{amsmath,amssymb,amsfonts}
\usepackage{algorithmic}
\usepackage{adjustbox}
\usepackage{graphicx}
\usepackage[para,online,flushleft]{threeparttable}
\usepackage{textcomp}
\usepackage[dvipsnames]{xcolor}
\usepackage{booktabs}
\usepackage{bm}
\def\BibTeX{{\rm B\kern-.05em{\sc i\kern-.025em b}\kern-.08em
    T\kern-.1667em\lower.7ex\hbox{E}\kern-.125emX}}
\begin{document}

\bstctlcite{BSTcontrol}

\title{FireFly-P: FPGA-Accelerated Spiking Neural Network Plasticity for Robust Adaptive Control}

\author{\normalsize
Tenglong Li\textsuperscript{\scriptsize *1,2,3,5} \enspace
Jindong Li\textsuperscript{\scriptsize *1,2,3,5} \enspace
Guobin Shen\textsuperscript{\scriptsize 1,2,3,4} \enspace
Dongcheng Zhao\textsuperscript{\scriptsize 2,3} \enspace
Qian Zhang\textsuperscript{\scriptsize 1,2,3,5} \enspace
Yi Zeng\textsuperscript{\scriptsize 1,2,3,4,5,6}\\
$^1$Brain-inspired Cognitive Intelligence Lab, Institute of Automation, Chinese Academy of Sciences\\ 
$^2$ Center for Long-term Artificial Intelligence,
$^3$ Beijing Institute of AI Safety and Governance\\
$^4$ School of Future Technology,
$^5$ School of Artificial Intelligence, University of Chinese Academy of Sciences  \\ 
$^6$ Key Laboratory of Brain Cognition and Brain-inspired Intelligence Technology, Chinese Academy of Sciences\\ 
{\{litenglong2023, lijindong2022, shenguobin2021,}{zhaodongcheng2016, q.zhang, yi.zeng\}@ia.ac.cn}
\thanks{*Equal Contribution. Corresponding author: Qian Zhang and Yi Zeng.}
}

\maketitle

\begin{abstract}
Spiking Neural Networks (SNNs) offer a biologically plausible learning mechanism through synaptic plasticity, enabling unsupervised adaptation without the computational overhead of backpropagation. To harness this capability for robotics, this paper presents FireFly-P, an FPGA-based hardware accelerator that implements a novel plasticity algorithm for real-time adaptive control. By leveraging on-chip plasticity, our architecture enhances the network's generalization, ensuring robust performance in dynamic and unstructured environments. The hardware design achieves an end-to-end latency of just 8~$\bm{\mu}$s for both inference and plasticity updates, enabling rapid adaptation to unseen scenarios. Implemented on a tiny Cmod A7-35T FPGA, FireFly-P consumes only 0.713~W and $\sim$10K~LUTs, making it ideal for power- and resource-constrained embedded robotic platforms. This work demonstrates that hardware-accelerated SNN plasticity is a viable path toward enabling adaptive, low-latency, and energy-efficient control systems.
\end{abstract}


\section{Introduction}
Synaptic plasticity forms the biological foundation of learning and memory by dynamically reorganizing neural connections in response to environmental changes~\cite{hebb2005organization, turrigiano1999homeostatic, abraham1996metaplasticity, abbott2000synaptic, bienenstock1982theory}. While essential for flexible behavior in neuroscience~\cite{bi1998synaptic}, translating this adaptability into artificial systems remains challenging. Direct application of plasticity often lacks the efficiency and precision required for complex computational tasks.

Harnessing this unsupervised adaptability is crucial for autonomous agents operating in unpredictable environments~\cite{kheradpisheh2016bio, lobov2020spatial, zalama1995real, park2021collision}. Unlike conventional offline-trained controllers that struggle with morphological changes or novel dynamics, adaptive plasticity enables online adjustment and recovery, maintaining functionality under disturbance~\cite{hajizada2022interactive, glatz2019adaptive}.

However, standard plasticity rules like Long-Term Potentiation (LTP)~\cite{lynch2004long}, Long-Term Depression (LTD)~\cite{ito1989long}, and Spike-Timing-Dependent Plasticity (STDP)~\cite{caporale2008spike, pfister2006triplets, lammie2018efficient, srinivasan2016magnetic, pfister2005beyond} are primarily limited to simple recognition tasks~\cite{diehl2015unsupervised, panda2016unsupervised}. Furthermore, existing neuromorphic hardware~\cite{davies2018loihi, pei2019towards, li2023firefly,li2024firefly, panchapakesan2022syncnn, aung2023deepfire2} offers limited support for plasticity models. The lack of efficient algorithms and hardware architectures supporting real-time adaptation continues to hinder practical robotic deployment.

\begin{figure}[t]
    \centering
    \includegraphics[width=\columnwidth]{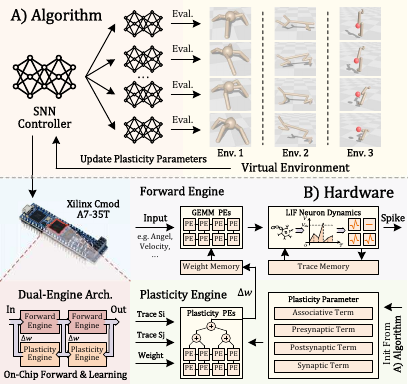}
    \caption{The FireFly-P framework pipeline. Plasticity parameters are first optimized in an offline stage. Subsequently, the learned rule is deployed on a dual-engine FPGA accelerator for online, real-time adaptive control.}
    \label{fig:introduction}
\end{figure}

In this work, we present FireFly-P, an algorithm-hardware co-design framework that brings SNN plasticity to robust and adaptive robotic control, as illustrated in Fig.~\ref{fig:introduction}. The main contributions of this paper are summarized as follows:

(1) We introduce a two-phase synaptic plasticity mechanism, as shown in Fig.~\ref{fig:introduction}A. The initial offline phase involves optimizing a plasticity rule on a representative task using an Evolutionary Strategy. Then, this learned rule is deployed online to continuously adjust synaptic weights, enabling the controller to adapt to dynamic and unforeseen task conditions.

(2) We design an efficient FPGA accelerator to implement the online plasticity process. The architecture includes a forward engine for low-latency SNN inference and a plasticity engine for real-time synaptic weight adjustment, as depicted in Fig.~\ref{fig:introduction}B. The proposed design achieves high performance within the limited resource and power budgets of embedded FPGA platforms, facilitating on-board deployment.

(3) We validate our system on a suite of challenging continuous control tasks, including direction, velocity, and position generalization, demonstrating robust adaptability in dynamic environments. Our hardware implementation achieves an 8~$\mu$s end-to-end latency and consumes just $\sim$10K~LUTs in 0.713~W, proving its suitability for real-world robotic applications.

\section{Algorithmic Framework}
The framework is built upon a synaptic plasticity mechanism described by a parametric update rule with coefficients optimized offline for continuous online adaptation, allowing the SNN to adjust synaptic weights in real time.

\subsection{Adaptive Synaptic Update Rule}
We introduce a parametric plasticity model in which the change in synaptic weight, $\Delta w_{ij}$, from a presynaptic neuron $j$ to a postsynaptic neuron $i$, is expressed as the sum of four functionally distinct components:
$$
\Delta w_{ij} = \underbrace{\alpha_{ij} S_j(t) S_i(t)}_{\text{Associative}} + \underbrace{\beta_{ij} S_j(t)}_{\text{Presynaptic}} + \underbrace{\gamma_{ij} S_i(t)}_{\text{Postsynaptic}} + \underbrace{\delta_{ij}}_{\text{Synaptic}}
$$
Here, $S_j$ and $S_i$ represent the spike traces of the pre- and postsynaptic neurons, respectively. A spike trace~\cite{pfister2006triplets} serves as an exponentially decaying memory of a neuron's recent firing activity, and is updated online according to:
$$
S(t) = \lambda S(t-1) + s(t)
$$
where $s(t) \in \{0, 1\}$ is a binary variable indicating a spike event at time $t$ (e.g., generated by the widely adopted Leaky Integrate-and-Fire neuron model~\cite{dayan2005theoretical, li2024fireflys, li2025fireflyt}), and $\lambda$ is a decay constant controlling the memory timescale.

Each component of the update rule, characterized by a learnable coefficient, serves a specific function:

\begin{itemize}
    \item \textbf{Associative Potentiation:} The Hebbian term ($\alpha_{ij}$)~\cite{hebb2005organization, duan2023hebbian} that strengthens synaptic connections when pre- and postsynaptic activities are temporally correlated, reinforcing causal relationships between neurons.
    
    \item \textbf{Presynaptic Depression:} The presynaptic term ($\beta_{ij}$) that weakens the synapse based solely on presynaptic activity, discouraging non-causal firing patterns.

    \item \textbf{Postsynaptic Homeostasis:} The postsynaptic term ($\gamma_{ij}$) that provides a homeostatic mechanism according to postsynaptic activity, supporting stable firing dynamics.

    \item \textbf{Synaptic Regularization:} The decay term ($\delta_{ij}$) that imposes an activity-independent, gradual reduction of synaptic strength, preventing unbounded growth and contributing to long-term stability of the network.
\end{itemize}

Collectively, the set of parameters $\theta = \{\alpha, \beta, \gamma, \delta\}$ defines the complete plasticity rule for a given synapse, which is learned through an offline optimization procedure as follows.

\subsection{Two-Phase Learning Framework}
Our strategy separates learning into two phases: offline rule optimization and online synaptic adaptation. This allows for a computationally intensive search for a robust rule that can then be deployed efficiently for real-time control.

\textbf{Phase 1: Offline Rule Optimization.} In this initial phase, the parameter space is explored using an Evolutionary Strategy (ES) to determine the optimal set of plasticity coefficients, $\theta^*$. A population of SNNs, each configured with a candidate parameter set, is evaluated on a representative task. Through iterative selection and mutation, the ES converges on a parameter set $\theta^*$ that produces robust adaptive behavior. The outcome of this phase is not a fine-tuned set of synaptic weights, but rather a \textbf{generalized learning rule.}

\textbf{Phase 2: Online Synaptic Adaptation.} With the optimized rule parameters $\theta^*$ frozen, the SNN is deployed on the target hardware (e.g., the robot's onboard FPGA). Starting from a zero-initialized state, its synaptic weights $w_{ij}$ are continuously updated by the learned plasticity rule during operation. This mechanism enables the SNN to autonomously reorganize its synaptic connections and develop compensatory behaviors in response to perturbations, such as simulated leg failure.

\section{Hardware Architecture}
Implementing online adaptation in Phase 2 requires hardware capable of rapid SNN inference and parallel synaptic updates with minimal latency. To this end, we propose a specialized hardware accelerator designed to meet these demands.

\begin{figure*}[t]
    \centering
    \includegraphics[width=\textwidth]{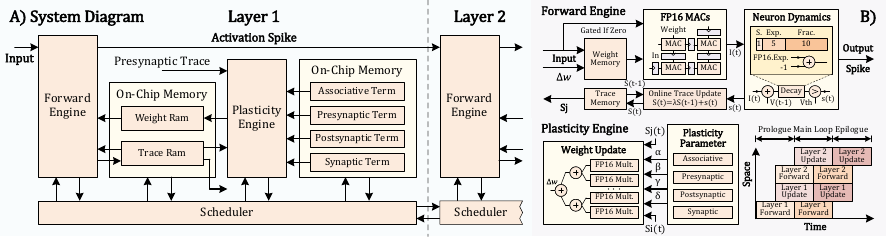}
    \caption{The FireFly-P hardware architecture, featuring a Dual-Engine Computation Core (Forward Engine for inference, Plasticity Engine for weight updates), a Scheduler, and an On-Chip Memory System, where multi-level pipelining is used to hide latency between the engines and across layers.}
    \label{fig:hardware}
\end{figure*}

\subsection{Overall Architecture}
As illustrated in Fig.~\ref{fig:hardware}A, the FireFly-P architecture integrates inference and plasticity learning within a unified FPGA pipeline using a tiling-based mapping strategy. It comprises four main modules: a \textbf{\textit{Forward Engine}}, a \textbf{\textit{Plasticity Engine}}, an \textbf{\textit{On-Chip Memory System}}, and a lightweight \textbf{\textit{Scheduler}}, cooperating to balance latency and hardware utilization.

The Scheduler coordinates the Forward Engine, responsible for SNN inference, and the Plasticity Engine, which handles synaptic updates. These engines operate in a tightly coupled, overlapping pipeline over a unified data path and memory interface. All computations employ 16-bit floating-point (FP16) arithmetic to balance sensitivity to small weight changes with resource efficiency. The On-Chip Memory System, built with BRAMs, stores all necessary network states and parameters, supporting tiled neuron mapping via strided addressing. To ensure data consistency, the Scheduler prioritizes memory accesses between the engines and manages their execution flow for precise, deterministic control.

\subsection{Dual-Engine Computation Core}
The central component of the FireFly-P accelerator is its Dual-Engine Computation Core, consisting of a Forward Engine for inference and a Plasticity Engine for synaptic adaptation, as illustrated in Fig.~\ref{fig:hardware}B. This specialized design maximizes each engine's performance, as detailed below.

\textbf{Forward Engine:} The Forward Engine is architected as a three-stage pipeline for each layer: a Psum Calculation Unit, a Neuron Dynamic Unit, and a Trace Update Unit. The first stage employs a partial sum (psum)-stationary dataflow, accumulating input current in local Processing Element (PE) registers to minimize on-chip memory access. Upon completion, the Neuron Dynamic Unit updates the membrane potential $V(t)$ based on the accumulated input current $I(t)$ using the Leaky Integrate-and-Fire (LIF) model:
$$
V(t) = V(t-1) + \frac{1}{{\tau_m}}(I(t)-V(t-1))
$$
Here, the membrane time constant $\tau_m$ is set to 2, which enables a multiplier-free implementation using only simple adders. When the membrane potential $V(t)$ exceeds a threshold $V_{th}$, a binary spike $s(t)$ is generated and broadcast to the next layer and the Trace Update Unit, which can be leveraged to gate downstream logic for dynamic power reduction.

\textbf{Plasticity Engine:} The Plasticity Engine is designed for the high-throughput computation of the four-component synaptic update rule. Its key architectural feature is a highly parallelized datapath. To maximize memory bandwidth, the four plasticity parameters $\{\alpha, \beta, \gamma, \delta\}$ for each synapse are packed and fetched in a single, wide memory access, as depicted in Fig.~\ref{fig:hardware}B. These parameters, along with the corresponding pre- and postsynaptic spike traces, are fed into an array of dedicated DSP blocks that compute all four terms of the plasticity rule concurrently. The results are then aggregated by a pipelined adder tree to produce the final weight change, $\Delta w_{ij}$.

\textbf{Scheduler:} The Scheduler ensures data coherency between the Forward and Plasticity Engines. It manages valid-data state transitions and arbitrates memory requests to the shared, dual-port BRAM, thus avoiding the need for double buffering. For example, the Forward Engine needs the latest weights from the Plasticity Engine, which can cause Read-After-Write (RAW) conflicts. To prevent this, a write-priority memory scheme pauses reads during writes, ensuring Forward Engine always uses up-to-date weights without introducing significant pipeline stalls. A similar arbitration mechanism is also applied to the trace memory to ensure deadlock-free data streaming.

\subsection{Pipeline Dataflow and Scheduling}
To maximize the engine's throughput, we designed a deeply pipelined dataflow, as shown in Fig.~\ref{fig:hardware}B, which overlaps the forward inference of one layer with the synaptic update of the preceding layer. For a two-layer SNN, a Scheduler orchestrates this execution through three distinct phases:

\textbf{Prologue:} The process begins with a forward pass of the first layer (L1). This computes the initial spike outputs from L1, which serve as the input for the second layer (L2).

\textbf{Main Loop:} Following initialization, the system enters a steady-state main loop for real-time operation. Each iteration proceeds in two main phases:
\begin{itemize}
    \item \textbf{Phase A: L1 Synaptic Update and L2 Forward Pass.} In this phase, L1 performs synaptic updates based on the spike traces from current timestep, while L2 simultaneously computes the forward pass using spike data produced by L1. This overlap effectively hides the latency of L1’s weight updates behind L2’s forward inference.

    \item \textbf{Phase B: L2 Synaptic Update and L1 Forward Pass.} In the subsequent phase, L2 updates its synaptic weights using the stable neuronal activities from the just-completed forward pass of L2. Concurrently, L1 receives the new input for the next timestep and computes the forward pass, preparing inputs for L2 in the following iteration.
\end{itemize}

\textbf{Epilogue:} Upon completion of the main loop, a final synaptic update is applied to L2 to incorporate the last set of neuronal activities and ensure all weights are current.

Notably, this phased description is simplified for clarity. In practice, forward and update operations within a layer can execute in parallel when data dependencies permit. This multi-level pipelining strategy—creating parallelism within engines, between them, and across layers—is what enables FireFly-P to support high-throughput, continuous and real-time adaptation.

\section{Implementation and Experiments}
\subsection{Experiments Setup}
We validated the FireFly-P framework through software simulation and hardware implementation: (1) optimizing the SNN controller's plasticity parameters, and (2) evaluating the hardware accelerator for on-chip learning and inference.

\textbf{Algorithm:} We use Parameter-Exploring Policy Gradients (PEPG)~\cite{sehnke2010parameter} to train three-layer fully connected SNN controllers. The architecture uses 128 hidden neurons for continuous control tasks (1024 for a separate MNIST task). Training is performed in three environments from the Brax simulator~\cite{freeman2021brax}: \textbf{ant}, trained to move toward 8 target directions and evaluated on generalization to 72 novel directions; \textbf{half cheetah}, trained to achieve 8 target velocities and tested on 72 unseen velocities; and \textbf{ur5e}, trained on a reaching task with randomly sampled goal positions.

\textbf{Hardware:} The FireFly-P accelerator was implemented on a Xilinx Artix-7 FPGA (Cmod A7-35T) with 16 PEs for Dual-Engine. The hardware design was described in SpinalHDL and then synthesized and implemented using Xilinx Vivado 2024.2, targeting a 200~MHz clock. All power consumption and resource utilization metrics were derived from the post-implementation reports generated by the Vivado Design Suite.

\subsection{Results and Analysis}
\begin{figure}[t]
    \centering
    \includegraphics[width=\columnwidth]{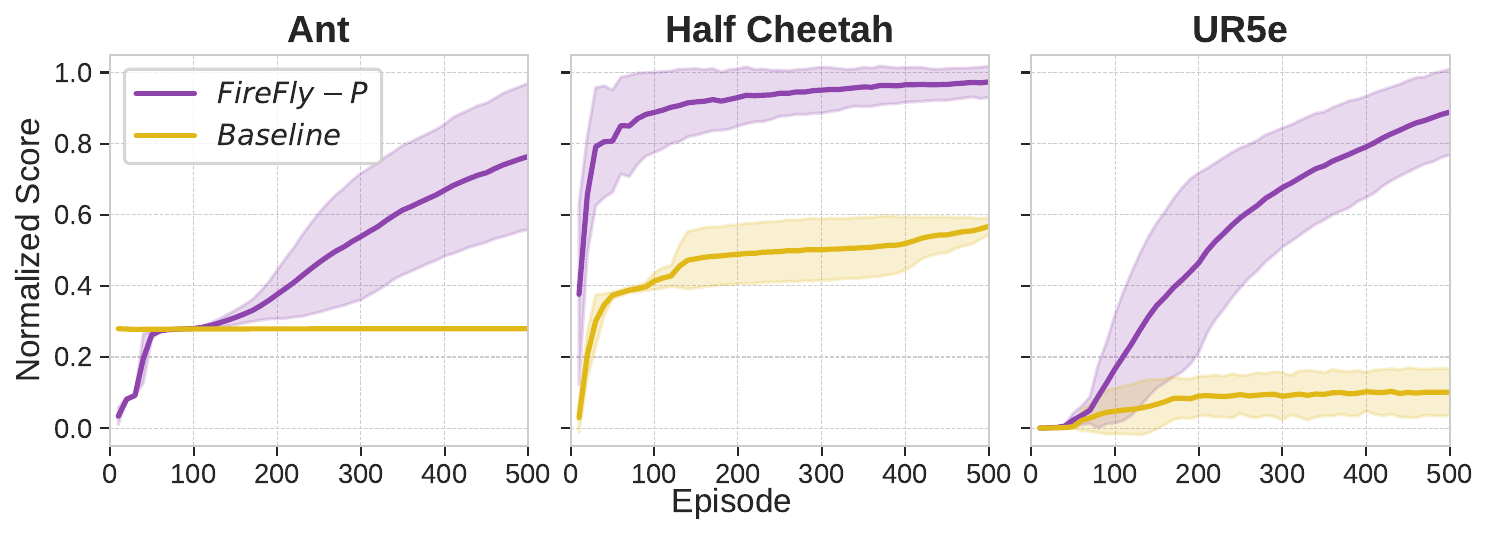}
    \caption{Performance of FireFly-P vs. weight-trained SNNs on three continuous control tasks, where FireFly-P adapts faster and achieves higher performance.}
    \label{fig:algorithm}
\end{figure}
\begin{figure}[t]
    \centering
    \includegraphics[width=\columnwidth]{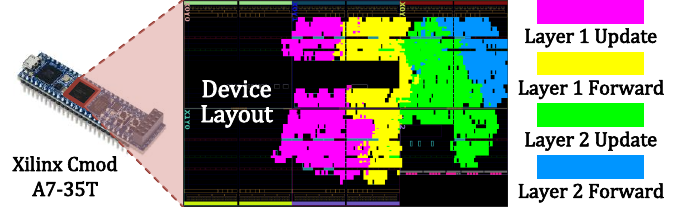}
    \caption{Implemented design layout of FireFly-P on a Xilinx Artix-7 FPGA.}
    \label{fig:layout}
\end{figure}
\textbf{Algorithm:} We evaluated FireFly-P against SNNs with directly trained synaptic weights on three reinforcement learning tasks. As shown in Fig.~\ref{fig:algorithm}, FireFly-P consistently demonstrated faster adaptation and superior overall performance. Agents using FireFly-P learned to generalize effectively to unseen target directions, velocities, and positions, whereas the weight-trained SNNs often converged to suboptimal or trivial behaviors. The smoother and more stable learning curves from FireFly-P indicate that its learned plasticity rules enable more flexible and efficient learning dynamics for continuous control.

\textbf{Hardware:} We validated the FireFly-P accelerator on real-time continuous control tasks to evaluate its suitability for edge robotics. As shown in Fig.~\ref{fig:layout} and Table~\ref{tab:resource}, the hardware design is highly compact, utilizing only 10.9k LUTs and 47 DSPs on a low-cost Cmod A7-35T FPGA, even with substantial logic dedicated to synaptic plasticity. This efficient co-design enables a pipelined architecture that completes a full inference-and-learning phase in just 8~$\mu$s while consuming only 0.713 W, confirming FireFly-P's capability to deliver powerful on-chip learning without compromising the real-time performance and low-power requirements of autonomous edge systems.
\begin{table}
  \setlength{\tabcolsep}{3.25pt}
  \centering
  \renewcommand\arraystretch{1.125}
  \caption{Resource Breakdown of FireFly-P for Continuous Control}
  \label{tab:resource}
  \begin{tabular}{ccccc}
    \toprule
    \textbf{Component} & \textbf{kLUTs} & \textbf{kREGs} & \textbf{BRAMs} & \textbf{DSPs} \\
    \midrule
    L1 Forward & 2.9 (14.10\%) & 3.5 (8.45\%) & 2 (4.00\%) & 12 (13.33\%) \\
    L1 Update & 3.1 (15.17\%) & 4.8 (11.57\%) & 0 (0.00\%) & 16 (17.78\%) \\
    L2 Forward & 1.6 (7.53\%) & 2.2 (5.26\%) & 0.5 (1.00\%) & 3 (3.33\%) \\
    L2 Update & 3.2 (15.14\%) & 4.8 (11.57\%) & 0 (0.00\%) & 16 (17.78\%) \\
    Others & 0.1 (0.88\%) & 1.3 (3.15\%) & 18 (36.00\%) & 0 (0.00\%) \\
    \midrule
    \textbf{Total} & \textbf{10.9 (52.82\%)} & \textbf{16.6(40.00\%)} & \textbf{20.5(41.00\%)} & \textbf{47(52.22\%)} \\
    \bottomrule
  \end{tabular}
\end{table}

Furthermore, to facilitate a fair comparison with prior works, we benchmarked the real-time performance of the FireFly-P accelerator on the MNIST classification task. As summarized in Table~\ref{tab:hardware}, FireFly-P attains a superior accuracy of 97.5\% with a compact SNN, outperforming previous hardware implementations. Notably, at a 200~MHz clock frequency, FireFly-P sustains an end-to-end throughput of 32 FPS, which integrates both inference and learning updates. This result surpasses prior systems that exhibit lower FPS due to their sequential execution of these stages.
\begin{table}[t]
\setlength{\tabcolsep}{4pt}
\centering
\renewcommand\arraystretch{1.125}
\begin{threeparttable}[b]
\caption{Performance of Edge-Computing SNN Hardware on MNIST}
\label{tab:hardware}
\begin{tabular}{cccccc}
\toprule
\textbf{Work} & \textbf{Learning Rule} & \textbf{Network} & \textbf{Acc.} & \textbf{FPS} & \textbf{Freq.}\\
\midrule
\cite{yousefzadeh2018practical} & Stochastic STDP & 784-6400-10 & 95.7 & - & 100 \\
\cite{li2021fast} & Pair-based STDP & 784-200-100-10 & 92.93 & 317 / 61 & 100 \\
\cite{neil2014minitaur} & Persistent CD & 784-500-500-10 & 92 & 1.89 / - & 75 \\
\cite{wang2017energy} & Pair-based STDP & 784-800 & 89.1 & 0.12 / 0.06 & 120 \\
\cite{ma2017darwin} & Persistent CD & 784-500-500-10 & 93.8 & 6.25 / - & 25 \\
\cite{he2021low} & Triplet R-STDP & 784-2048-100 & 93 & 30 / 22.5 & 200 \\
\textbf{Ours} & \textbf{Learnable STDP} & \textbf{784-1024-10} & \textbf{97.5} & \textbf{32\tnote{1}} & \textbf{200} \\
\bottomrule
\end{tabular}\begin{tablenotes}
    \item[1] A/B denotes forward/learning FPS. Our method pipelines these two stages, and we thus report the end-to-end FPS including both forward and learning.
\end{tablenotes}
\end{threeparttable}
\end{table}

\section{Conclusion}
We proposed FireFly-P, an algorithm-hardware co-design framework that brings real-time, adaptive control to robotic systems through SNN plasticity. By pairing a novel, offline-optimized plasticity rule with a highly efficient dual-engine FPGA accelerator, FireFly-P achieves continuous on-chip learning while consuming only 0.713~W and utilizing $\sim$10K LUTs. Experiments demonstrate rapid recovery and robust performance in dynamic tasks, highlighting its potential for autonomous, adaptive, and energy-efficient intelligent systems.

\section{Acknowledgment}
This work was supported by the General Program of the National Natural Science Foundation of China (Grant Nos. 32571284, 62406325, and 32441109), the State Key Laboratory of Brain Cognition and Brain-inspired Intelligence Technology (Grant No. JS202401), the Artificial Intelligence of General Anesthesia Mechanism Program of Yichang Humanwell (Grant No. Y8D1051LE1), and the Institute of Automation, Chinese Academy of Sciences (Grant No. E411230101).

\bibliographystyle{IEEEtran}
\bibliography{myIEEE,reference}

\begin{thebibliography}{10}
\providecommand{\url}[1]{#1}
\csname url@samestyle\endcsname
\providecommand{\newblock}{\relax}
\providecommand{\bibinfo}[2]{#2}
\providecommand{\BIBentrySTDinterwordspacing}{\spaceskip=0pt\relax}
\providecommand{\BIBentryALTinterwordstretchfactor}{4}
\providecommand{\BIBentryALTinterwordspacing}{\spaceskip=\fontdimen2\font plus
\BIBentryALTinterwordstretchfactor\fontdimen3\font minus \fontdimen4\font\relax}
\providecommand{\BIBforeignlanguage}[2]{{%
\expandafter\ifx\csname l@#1\endcsname\relax
\typeout{** WARNING: IEEEtran.bst: No hyphenation pattern has been}%
\typeout{** loaded for the language `#1'. Using the pattern for}%
\typeout{** the default language instead.}%
\else
\language=\csname l@#1\endcsname
\fi
#2}}
\providecommand{\BIBdecl}{\relax}
\BIBdecl

\bibitem{hebb2005organization}
Donald~Olding Hebb, \emph{The organization of behavior: A neuropsychological theory}.\hskip 1em plus 0.5em minus 0.4em\relax Psychology press, 2005.

\bibitem{turrigiano1999homeostatic}
Gina~G Turrigiano, ``Homeostatic plasticity in neuronal networks: the more things change, the more they stay the same,'' \emph{Trends in neurosciences}, vol.~22, no.~5, pp. 221--227, 1999.

\bibitem{abraham1996metaplasticity}
Wickliffe~C Abraham and Mark~F Bear, ``Metaplasticity: the plasticity of synaptic plasticity,'' \emph{Trends in neurosciences}, vol.~19, no.~4, pp. 126--130, 1996.

\bibitem{abbott2000synaptic}
Larry~F Abbott and Sacha~B Nelson, ``Synaptic plasticity: taming the beast,'' \emph{Nature neuroscience}, vol.~3, no.~11, pp. 1178--1183, 2000.

\bibitem{bienenstock1982theory}
Elie~L Bienenstock, Leon~N Cooper, and Paul~W Munro, ``Theory for the development of neuron selectivity: orientation specificity and binocular interaction in visual cortex,'' \emph{Journal of Neuroscience}, vol.~2, no.~1, pp. 32--48, 1982.

\bibitem{bi1998synaptic}
Guo-qiang Bi and Mu-ming Poo, ``Synaptic modifications in cultured hippocampal neurons: dependence on spike timing, synaptic strength, and postsynaptic cell type,'' \emph{Journal of neuroscience}, vol.~18, no.~24, pp. 10\,464--10\,472, 1998.

\bibitem{kheradpisheh2016bio}
Saeed~Reza Kheradpisheh, Mohammad Ganjtabesh, and Timoth{\'e}e Masquelier, ``Bio-inspired unsupervised learning of visual features leads to robust invariant object recognition,'' \emph{Neurocomputing}, vol. 205, pp. 382--392, 2016.

\bibitem{lobov2020spatial}
Sergey~A Lobov, Alexey~N Mikhaylov, Maxim Shamshin, Valeri~A Makarov, and Victor~B Kazantsev, ``Spatial properties of stdp in a self-learning spiking neural network enable controlling a mobile robot,'' \emph{Frontiers in neuroscience}, vol.~14, p.~88, 2020.

\bibitem{zalama1995real}
Eduardo Zalama, Paolo Gaudiano, and Juan~L{\'o}pez Coronado, ``A real-time, unsupervised neural network for the low-level control of a mobile robot in a nonstationary environment,'' \emph{Neural networks}, vol.~8, no.~1, pp. 103--123, 1995.

\bibitem{park2021collision}
Kyu~Min Park, Younghyo Park, Sangwoong Yoon, and Frank~C Park, ``Collision detection for robot manipulators using unsupervised anomaly detection algorithms,'' \emph{IEEE/ASME Transactions on Mechatronics}, vol.~27, no.~5, pp. 2841--2851, 2021.

\bibitem{hajizada2022interactive}
Elvin Hajizada, Patrick Berggold, Massimiliano Iacono, Arren Glover, and Yulia Sandamirskaya, ``Interactive continual learning for robots: a neuromorphic approach,'' in \emph{Proceedings of the International Conference on Neuromorphic Systems 2022}, 2022, pp. 1--10.

\bibitem{glatz2019adaptive}
Sebastian Glatz, Julien Martel, Raphaela Kreiser, Ning Qiao, and Yulia Sandamirskaya, ``Adaptive motor control and learning in a spiking neural network realised on a mixed-signal neuromorphic processor,'' in \emph{2019 international conference on robotics and automation (ICRA)}.\hskip 1em plus 0.5em minus 0.4em\relax IEEE, 2019, pp. 9631--9637.

\bibitem{lynch2004long}
Marina~A Lynch, ``Long-term potentiation and memory,'' \emph{Physiological reviews}, 2004.

\bibitem{ito1989long}
Masao Ito, ``Long-term depression.'' \emph{Annual review of neuroscience}, vol.~12, pp. 85--102, 1989.

\bibitem{caporale2008spike}
Natalia Caporale and Yang Dan, ``Spike timing--dependent plasticity: a hebbian learning rule,'' \emph{Annu. Rev. Neurosci.}, vol.~31, no.~1, pp. 25--46, 2008.

\bibitem{pfister2006triplets}
Jean-Pascal Pfister and Wulfram Gerstner, ``Triplets of spikes in a model of spike timing-dependent plasticity,'' \emph{Journal of Neuroscience}, vol.~26, no.~38, pp. 9673--9682, 2006.

\bibitem{lammie2018efficient}
Corey Lammie, Tara~Julia Hamilton, Andr{\'e} Van~Schaik, and Mostafa~Rahimi Azghadi, ``Efficient fpga implementations of pair and triplet-based stdp for neuromorphic architectures,'' \emph{IEEE Transactions on Circuits and Systems I: Regular Papers}, vol.~66, no.~4, pp. 1558--1570, 2018.

\bibitem{srinivasan2016magnetic}
Gopalakrishnan Srinivasan, Abhronil Sengupta, and Kaushik Roy, ``Magnetic tunnel junction based long-term short-term stochastic synapse for a spiking neural network with on-chip stdp learning,'' \emph{Scientific reports}, vol.~6, no.~1, p. 29545, 2016.

\bibitem{pfister2005beyond}
Jean-Pascal Pfister and Wulfram Gerstner, ``Beyond pair-based stdp: A phenomenological rule for spike triplet and frequency effects,'' \emph{Advances in neural information processing systems}, vol.~18, 2005.

\bibitem{diehl2015unsupervised}
Peter~U Diehl and Matthew Cook, ``Unsupervised learning of digit recognition using spike-timing-dependent plasticity,'' \emph{Frontiers in computational neuroscience}, vol.~9, p.~99, 2015.

\bibitem{panda2016unsupervised}
Priyadarshini Panda and Kaushik Roy, ``Unsupervised regenerative learning of hierarchical features in spiking deep networks for object recognition,'' in \emph{2016 international joint conference on neural networks (IJCNN)}.\hskip 1em plus 0.5em minus 0.4em\relax IEEE, 2016, pp. 299--306.

\bibitem{davies2018loihi}
Mike Davies, Narayan Srinivasa, Tsung-Han Lin, Gautham Chinya, Yongqiang Cao, Sri~Harsha Choday, Georgios Dimou, Prasad Joshi, Nabil Imam, Shweta Jain \emph{et~al.}, ``Loihi: A neuromorphic manycore processor with on-chip learning,'' \emph{Ieee Micro}, vol.~38, no.~1, pp. 82--99, 2018.

\bibitem{pei2019towards}
Jing Pei, Lei Deng, Sen Song, Mingguo Zhao, Youhui Zhang, Shuang Wu, Guanrui Wang, Zhe Zou, Zhenzhi Wu, Wei He \emph{et~al.}, ``Towards artificial general intelligence with hybrid tianjic chip architecture,'' \emph{Nature}, vol. 572, no. 7767, pp. 106--111, 2019.

\bibitem{li2023firefly}
Jindong Li, Guobin Shen, Dongcheng Zhao, Qian Zhang, and Yi~Zeng, ``Firefly: A high-throughput hardware accelerator for spiking neural networks with efficient dsp and memory optimization,'' \emph{IEEE Transactions on Very Large Scale Integration (VLSI) Systems}, vol.~31, no.~8, pp. 1178--1191, 2023.

\bibitem{li2024firefly}
------, ``Firefly v2: Advancing hardware support for high-performance spiking neural network with a spatiotemporal fpga accelerator,'' \emph{IEEE Transactions on Computer-Aided Design of Integrated Circuits and Systems}, vol.~43, no.~9, pp. 2647--2660, 2024.

\bibitem{panchapakesan2022syncnn}
Sathish Panchapakesan, Zhenman Fang, and Jian Li, ``Syncnn: Evaluating and accelerating spiking neural networks on fpgas,'' \emph{ACM Transactions on Reconfigurable Technology and Systems}, vol.~15, no.~4, pp. 1--27, 2022.

\bibitem{aung2023deepfire2}
Myat Thu~Linn Aung, Daniel Gerlinghoff, Chuping Qu, Liwei Yang, Tian Huang, Rick Siow~Mong Goh, Tao Luo, and Weng-Fai Wong, ``Deepfire2: A convolutional spiking neural network accelerator on fpgas,'' \emph{IEEE Transactions on Computers}, vol.~72, no.~10, pp. 2847--2857, 2023.

\bibitem{dayan2005theoretical}
Peter Dayan and Laurence~F Abbott, \emph{Theoretical neuroscience: computational and mathematical modeling of neural systems}.\hskip 1em plus 0.5em minus 0.4em\relax MIT press, 2005.

\bibitem{li2024fireflys}
Tenglong Li, Jindong Li, Guobin Shen, Dongcheng Zhao, Qian Zhang, and Yi~Zeng, ``Firefly-s: Exploiting dual-side sparsity for spiking neural networks acceleration with reconfigurable spatial architecture,'' \emph{IEEE Transactions on Circuits and Systems I: Regular Papers}, 2024.

\bibitem{li2025fireflyt}
------, ``Firefly-t: High-throughput sparsity exploitation for spiking transformer acceleration with dual-engine overlay architecture,'' \emph{arXiv preprint arXiv:2505.12771}, 2025.

\bibitem{duan2023hebbian}
Yu~Duan, Zhongfan Jia, Qian Li, Yi~Zhong, and Kaisheng Ma, ``Hebbian and gradient-based plasticity enables robust memory and rapid learning in rnns,'' \emph{arXiv preprint arXiv:2302.03235}, 2023.

\bibitem{sehnke2010parameter}
Frank Sehnke, Christian Osendorfer, Thomas R{\"u}ckstie{\ss}, Alex Graves, Jan Peters, and J{\"u}rgen Schmidhuber, ``Parameter-exploring policy gradients,'' \emph{Neural Networks}, vol.~23, no.~4, pp. 551--559, 2010.

\bibitem{freeman2021brax}
C~Daniel Freeman, Erik Frey, Anton Raichuk, Sertan Girgin, Igor Mordatch, and Olivier Bachem, ``Brax--a differentiable physics engine for large scale rigid body simulation,'' \emph{arXiv preprint arXiv:2106.13281}, 2021.

\bibitem{yousefzadeh2018practical}
Amirreza Yousefzadeh, Evangelos Stromatias, Miguel Soto, Teresa Serrano-Gotarredona, and Bernab{\'e} Linares-Barranco, ``On practical issues for stochastic stdp hardware with 1-bit synaptic weights,'' \emph{Frontiers in neuroscience}, vol.~12, p. 665, 2018.

\bibitem{li2021fast}
Sixu Li, Zhaomin Zhang, Ruixin Mao, Jianbiao Xiao, Liang Chang, and Jun Zhou, ``A fast and energy-efficient snn processor with adaptive clock/event-driven computation scheme and online learning,'' \emph{IEEE Transactions on Circuits and Systems I: Regular Papers}, vol.~68, no.~4, pp. 1543--1552, 2021.

\bibitem{neil2014minitaur}
Daniel Neil and Shih-Chii Liu, ``Minitaur, an event-driven fpga-based spiking network accelerator,'' \emph{IEEE transactions on very large scale integration (VLSI) systems}, vol.~22, no.~12, pp. 2621--2628, 2014.

\bibitem{wang2017energy}
Qian Wang, Youjie Li, Botang Shao, Siddhartha Dey, and Peng Li, ``Energy efficient parallel neuromorphic architectures with approximate arithmetic on fpga,'' \emph{Neurocomputing}, vol. 221, pp. 146--158, 2017.

\bibitem{ma2017darwin}
De~Ma, Juncheng Shen, Zonghua Gu, Ming Zhang, Xiaolei Zhu, Xiaoqiang Xu, Qi~Xu, Yangjing Shen, and Gang Pan, ``Darwin: A neuromorphic hardware co-processor based on spiking neural networks,'' \emph{Journal of systems architecture}, vol.~77, pp. 43--51, 2017.

\bibitem{he2021low}
Zhen He, Cong Shi, Tengxiao Wang, Ying Wang, Min Tian, Xichuan Zhou, Ping Li, Liyuan Liu, Nanjian Wu, and Gang Luo, ``A low-cost fpga implementation of spiking extreme learning machine with on-chip reward-modulated stdp learning,'' \emph{IEEE Transactions on Circuits and Systems II: Express Briefs}, vol.~69, no.~3, pp. 1657--1661, 2021.

\end{thebibliography}

\end{document}